# Five Questions on Complexity

Francis Heylighen
*ECCO, Vrije Universiteit Brussel*

### 1. Why did you begin working with complex systems?

I have been interested by all forms of complexity and self-organization since my childhood. I was always a keen observer of nature, being fascinated by complex phenomena such as ants walking apparently randomly across a branch, the cracks that would appear in drying mud, or the frost crystals that would form on grass during winter nights.

As an adolescent, one of my hobbies was keeping aquariums, in which I would try to build a miniature ecosystem complete with soil, plants, invertebrates, and fish. The fish would still need to get food from time to time, and I still had to clean the filter that would collect the dirt they produced, but ideally I would have liked to create a system that is completely autonomous, and is able to sustain itself even in the absence of a caretaker. That would have required more plant life to sustain the food chain, and especially less fish to produce waste products, so it would have made the aquarium less interesting to look at. Therefore, I did compromise in practice. But in my imagination, I was fascinated by what I called "a little world on its own". In my present scientific vocabulary, I would define this idea as a system that is complex and self-organizing to such a degree that it could be viewed almost as a separate, autonomous universe. (Later I discovered a similar idea in the science fiction stories of Stanislaw Lem, a Polish author influenced by cybernetics.)

My fascination for rocks, plants, animals and other phenomena of nature also found an outlet in my early inquiry into the theory of evolution. Like most children nowadays, I had been exposed from an early age to pictures and stories about dinosaurs. The difference, perhaps, is that my grandfather who had collected or drawn these pictures for me was rather scientifically minded, although he was just a primary school teacher. He taught me not only their Latin names, such as Brontosaurus, Triceratops and Tyrannosaurus Rex, but also about the periods in which they lived, and the kinds of creatures that preceded and followed them in the course of natural history. So, from an age of eight or so, I was well aware that life on Earth had evolved, and that plants and animals looked very different in different time periods.

As I became a little older, I started reading introductory books on biology, which explained the mechanism of natural selection behind this evolution. This idea became one of the two fundamental principles on which I have based my scientific worldview. As an adolescent, this mechanism seemed so obvious to me that I was quick to generalize it to other domains, noting that for example ideas and societies also evolved through variation and selection. I called this "the generalized principle of natural selection". Much later, while working on my PhD, I came into contact with other scientists (in particular the great Donald T. Campbell and his disciples Gary Cziko and Mark Bickhard) who had developed a similar philosophy, which they called "selectionism" or "universal selection theory". Its basic assumption is that all complex

systems—whether physical, biological, mental or social—have originated through an evolutionary process, which at the deepest level consists of some form of "blind" (not necessarily random) variation, followed by the selective retention of those variants that are most "fit".

In this radical formulation, the theory has few adherents. The reason is that most complexity scientists view Darwin's theory of natural selection with its emphasis on individual organisms or genes as reductionist, ignoring the "whole is greater than the sum of the parts" mantra that characterizes self-organization and complex systems. Yet, I never saw a contradiction between this holistic perspective and my beloved principle of natural selection. The explanation lies in another fundamental idea that I developed while I was 15-16 years old, and which I called the "relational principle".

After reading popular science accounts of Einstein's theory of relativity, I was inclined to conclude, like so many others with a somewhat rebellious streak, that "everything is relative", and that there are no absolute laws nor truths, neither man-made nor natural. (Later I learned that Einstein's own philosophy could hardly have been more different). More recently, this irreverent philosophy has gotten some form of academic respectability under the label of "postmodernism" or "social constructivism". Its main thesis is that different cultures and different people see the same things in different ways, and that there is no absolute criterion to say who is right and who is wrong. But this negative interpretation did not satisfy me: I wanted to truly understand how the world functions.

Therefore, I focused on the positive aspect of relativity: the importance of relations. A phenomenon can only be conceived with respect to, or *in relation to*, another phenomenon. No phenomenon can exist on its own—without context or environment from which it is distinguished, but to which it is at the same time connected. Later, this idea led me to analyze everything in terms of "bootstrapping" networks, where nodes are defined by their links with other nodes, and links by the nodes they connect. This philosophy is intrinsically holistic: it is impossible to reduce systems to their separate components; it is only through the connections between the components that the system emerges. This relational point of view is not in conflict with selectionism: networks do undergo variation and selection, both at the level of the nodes and links that constitute them and at the higher level of the systems that emerge from clusters of densely linked nodes.

After having formulated the fundamental tenets of my philosophy already while in high school, my challenge was to choose a discipline to study in university. With such a broad interest in complex systems of all types (I had even "reinvented" the concept of social network by drawing a map of all the relationships within my high school class—an exercise that did not make me too popular among my classmates), coupled with a healthy skepticism towards traditional reductionist science, this was not an obvious issue. I hesitated between biology, physics, philosophy and literature, and finally settled on physics, reasoning that I could study the other ones on my own, but with the math underlying physics being so difficult, I would need some solid tutoring if I wanted to become mathematically literate enough to understand the most advanced theories. This reasoning turned out to be correct: studying theoretical physics was hard, but it gave me a basis that allowed me to afterwards investigate a variety of other scientific disciplines on my own.

Within physics, my interest initially did not go towards complexity—which at the time (around 1980) was not yet a fashionable topic. I was lucky enough to get some courses on thermodynamics and statistical mechanics from professors who had worked with the great Ilya Prigogine, the founder of the Brussels School of complex systems. But these particular individuals were less inspiring to me than a young assistant researcher, Diederik Aerts, who was investigating the foundations of quantum mechanics. So, I decided to make, first my Master's thesis, then my PhD on that subject, hoping to be able to elaborate my relational philosophy in a more formal manner. An analysis of the role of the observer in quantum theory together with the creation at our university by Luc Steels of one of the first Artificial Intelligence labs in Europe inspired me to focus on cognition: the processes by which knowledge is acquired and represented.

I submitted a short paper looking at knowledge acquisition as relational self-organization to a conference on cybernetics. There I discovered a whole community of researchers interested in the same transdisciplinary subject of complex systems, their self-organization and cognition. After defending my PhD thesis in 1987, I basically abandoned my work on the foundations of physics, and positioned myself squarely in the field of general systems and cybernetics, hoping that I had finally found my home. Yet, I felt there was still something lacking in that approach, which tended to consider systems as pre-existing, static structures. I missed the evolutionary angle. Therefore I wrote a "Proposal for the creation of a network on complexity research", sketching a theoretical framework that would integrate systems, evolution and cognition.

From the reactions I received, the most interesting one came from a young systems scientist, Cliff Joslyn, who had just developed a similar proposal in collaboration with the veteran cyberneticist Valentin Turchin. They called it the "Principia Cybernetica Project". In 1991 I joined them, and in 1993 I created the project's website. Principia Cybernetica Web (http://pcp.vub.ac.be) was the first and still is one of the most important websites on complex systems, cybernetics, evolution, and related subjects. As such, it has gotten countless students and researchers interested in the domain.

Since then I have been working on integrating these different topics in an encompassing theoretical framework, with a variety of applications in social systems, information technology, psychology and related domains. Independently of our "evolutionary cybernetics" work in Principia Cybernetica, the complex adaptive systems approach had in the meantime become popular, thanks mostly to researchers affiliated with the Santa Fe Institute, such as John Holland and Stuart Kauffman. The similarities between both approaches are much more important than the differences, but there is still enough difference in focus to allow for useful cross-fertilization. It was in part for this purpose that in 2004 I founded the Evolution, Complexity and Cognition (ECCO) research group, which groups most of my PhD students and a number of associate researchers.

## 2. How would you define complexity?

Anticipating one of the following questions, arguably the most problematic aspect of complexity is its definition. Dozens if not hundreds of authors have proposed definitions, some vague and qualitative, some formal and quantitative, but none of them really satisfactory. The formal ones tend to be much too specific, being applicable only to binary strings or to genomes, but not to complex systems in general. Moreover, even within the extremely simplified universe of binary strings (sequences of 0s and 1s), complexity turns out to be tricky to define. The best definition yet defines the complexity of a string as the length of the shortest possible complete description of it (i.e. the binary program needed to generate the string). However, this implies that a random string would be maximally complex.

The qualitative descriptions can be short and vague, such as "complexity is situated in between order and disorder". More commonly, authors trying to characterize complex systems just provide extensive lists or tables of properties that complex systems have and that distinguish them from simple system. These include items such as: many components or agents, local interactions, non-linear dynamics, emergent properties, self-organization, multiple feedback loops, multiple levels, adapting to its environment, etc. The problem here of course is that the different lists partly overlap, partly differ, and that there is no agreement on what should be included. Moreover, the properties are usually not independent. For example, self-organizing processes normally produce emergent properties, and include feedback loops, which themselves entail non-linearity… Then, not all properties are truly necessary. For example, as I recently noted at a conference where one of such definitions was proposed, a marriage is typically a very complex system that is unpredictable, non-linear, adaptive, etc. Yet it consists of just two agents!

For my own preferred definition, I go back to the Latin root "complexus", which means something like "entangled, entwined, embracing". I interpret this to mean that in order to have a complex, you need two or more distinct components that are connected in such a way that they are difficult to separate. This fits in perfectly with my relational philosophy: it is the relations weaving the parts together that turn the system into a complex, producing emergent properties. To make this qualitative notion more quantitative, I add that a system becomes more complex as the number of distinctions (distinct components, states, or aspects) and the number of relations or connections increases.

The problem with this definition is that it does not lead to a unique number or degree that would allow us to objectively measure how complex a phenomenon is. The reason is that distinctions and connections are not objectively given, easily countable entities: they exist at different levels, in different dimensions, and in different kinds. Aspects can be related to each other across space, across time or across levels. Distinctions can be logical, physical, causal, or perceptual. Adding them all together in order to calculate the overall complexity of a system would be like adding apples and oranges. At best, this definition leads to what in mathematics is called a partial order: X might be more complex than Y, less complex, equally complex, or simply incomparable. It is more complex only if X has all the components and relations that Y has, plus some more.

In spite of this limitation, this definition has some nice characteristics: it is simple and intuitive, and it maps neatly on some of the other simple definitions. For example, complexity, characterized by many distinctions and connections, is situated in between disorder (many distinctions, few or no connections) and order (many connections, few or no distinctions). It also connects the relational and selectionist perspectives: an evolutionary process can be seen as a system of distinctions (variations) and connections (selective continuations) across time. Moreover, evolution generates complexity by increasing variety (number of distinct systems or states) and dependency (systems "fitting" or adapting to each other). I call these twin aspects of complexification: differentiation and integration.

## 3. What is your favourite aspect/concept of complexity?

As one might have guessed from my biographical notes, I am fascinated by self-organization. Unlike authors like Kauffman, I don't make a strict distinction between self-organization and evolution: both are processes that spontaneously take place in complex systems and that generate more complexity. Evolution tends to be seen in terms of adaptation to an external environment and self-organization as the result of an internal dynamics. Yet, from a systems perspective there is no absolute difference between internal and external: what is internal for the system is generally external for its subsystems. It all depends on where you draw the boundary between system and environment. Thus, as the cybernetician Ashby pointed out long ago, we can view any self-organizing system as a collection of co-evolving or mutually adapting subsystems. Similarly, we can view biological evolution as the self-organization of the ecosystem into a network of mutually adapted species.

I am not just interested in observing self-organization "in the wild", but in creating it in artificial systems. The best-known examples are the computer simulations of organisms, ecosystems and societies that we find in the domains of Artificial Life and Multi-Agent Systems. Such simulations have shown that very simple algorithms (abstract representations of iterative processes) can lead to unexpected complexity, adaptation, and apparently intelligent organization. Let's look at two classic examples.

Genetic algorithms are based on a simple generalization of Darwinian evolution. A variety of potential solutions to a particular problem are generated in the form of strings of symbols. These are tested as to their "fitness", or goodness in tackling the problem. The fittest candidates are selected and made to undergo variation, either by mutation (randomly changing one or a few symbols in the string) or by "sexual recombination" (gluing the first part of one string together with the last part of another). This produces a second generation, which is again selected on the basis of fitness. The best ones of the second generation then reproduce to form a third generation, and so on. After several such generations, the fittest string is typically much better than the ones you started out with, and often produces an elegant solution to a complex problem.

Ant algorithms too are directly inspired by natural self-organization: when ants find food, they leave a trail of pheromones ("smell molecules") along their path back to the nest. Other ants searching for food are more likely to go in a direction where there are more pheromones. If successful, they too will add pheromones, making the trail stronger, and more likely to attract further ants. If no food is found, no pheromones are added and the trail gradually evaporates. In that way, a colony of ants will at first explore their environment randomly, but then gradually develop a complex "roadmap" of trails connecting the nest and the various food sources in the most efficient way.

Applications of self-organization are found not only in computing or in nature, but also in society. Cities, communities, cultures and markets typically emerge through self-organization. Different people with different backgrounds meet by chance, exchange products, services or ideas, thus discovering common interests. This leads to an explicit or implicit collaboration, which is in everybody's interests, and thus binds the assembly of individuals together into a system. The system complexifies as people specialize in certain roles, thus creating a division a labor. This differentiation is counterbalanced by integration, through the creation of communication channels connecting the subsystems together into a larger whole. In that way, a hierarchy of levels is created. Eventually a single individual, such as a president, king, or mayor, may come to occupy the top level, apparently being in charge. But the system is much too complex to be centrally controlled: its "governor" (to use the cybernetic term) may specify high level goals and directions, but the concrete activities are still produced "bottom-up", through the interactions between individuals and subsystems.

Understanding this dynamics allows us to encourage and support it, e.g. when creating new social systems. This happens routinely on the Internet where virtual communities self-assemble around a website or discussion forum. I am particularly interested in the software tools that facilitate such self-organization, and have extensively researched the way they may enhance the "collective intelligence" of the emerging system. Such software tools typically support and guide the interaction between individuals and the information they use, recommending people or resources likely to be useful, and shortcutting the many trial-and-error processes that otherwise would be needed to find an adapted network, e.g. by using an equivalent of ant algorithms.

## 4. In your opinion, what is the most problematic aspect/concept of complexity?

Conceptually, the most difficult aspect of complexity is still its definition, and the deeper understanding that goes with it. This is probably because complexity requires us to abandon our traditional reductionist perspective, that is to say, our tendency to tackle complex systems by analyzing them into separate components. The opposite perspective of holism, on the other hand, runs the danger of too much vagueness and simplification: just noting that everything is connected to everything else is of little help when tackling concrete problems. The twin principles of relationalism and selectionism, as I sketched them, hold the promise of synthesizing these complementary approaches. Yet, they still remain quite abstract, and need to be developed into a more concrete and coherent theory.

Practically, the most problematic aspect of complexity is simply coping with it. It is hardly an original observation that our present society is getting more complex every day. The main reason is that modern communication and transport technologies have facilitated interactions between previously remote people, societies or systems, thus increasing their connectivity. Yet, I think that this phenomenon is still insufficiently studied. Indeed, many of our most pressing problems have this growing interdependency at their core.

Let me list some well-known example. Few people nowadays dispute the dangers of global warming. Yet, when it comes to tackling the problem, no one seems to know very well where to start: there are dozens of different possible strategies, from promoting alternative energy to instating a carbon tax, from planting more forests to injecting carbon dioxide into the soil… All of these have different disadvantages and costs attached to them, but—more importantly—they all interact, via their effect on the economy and the ecosystem. This makes the overall effect of any mix of strategies unpredictable. A recently "hot topic" in complexity science is the modeling and detection of terrorist networks. As the world becomes more interdependent, the potential damage created by terrorism grows, yet the terrorist groups become more diffuse and distributed, without a central command that is easy to take out. Finally, the explosive growth of the Internet has brought many benefits, but also created new problems, including information overload and the concomitant stress, cybercrime, and the spread of computer viruses and spam.

The only way we will be able to deal with such dynamic problems is to combat complexity with complexity, i.e. create models and systems based on the same principles of complexity and self-organization as the problem domains they are dealing with. As such, they can co-evolve with the problems, becoming ever better adapted to their moving targets. An illustration of such an approach can be found in the attempts to design a computer security system inspired by the mechanisms of our own immune system. This means that the system would learn to recognize and neutralize computer viruses, worms, intruders, and bugs by the variation and selection of "antibodies" that recognize and disable anything that doesn't behave as it should.

## 5. How do you see the future of complexity?

In the longer term, I see some form of complexity science take over the whole of scientific thinking, replacing the still lingering Newtonian paradigm, with its assumptions of separate components, predictable behavior, and static, unchanging laws. However, it is not obvious whether this will be the present, as yet poorly organized, incarnation of complexity science, or some future version that goes under a different name. As the critic John Horgan pointed out, the present complexity wave fits nicely in a sequence of "c-words" that became popular with intervals of about 15 years, but went out of fashion shortly afterwards: Cybernetics, Catastrophe theory, Chaos theory, and now Complexity.

Such ebb and flow of scientific fashions is certainly not limited to complexity. More important than the changes in focus and the accompanying buzzwords, however, is the continuity in the development of the underlying way of thinking. Most complexity researchers would agree that the basic ideas of cybernetics, catastrophe theory and chaos theory still nicely fit under the broad umbrella of complexity science. It is just that we have learned that very specific, and especially mathematical models, such as catastrophes, chaos, fractals, or more recently self-organized criticality, are useful only in a particular, well-defined context, and will need to be complemented by other approaches if we want to apply them to complex systems in general.

The danger is that complexity science would merely become an assortment of advanced modeling techniques that capture with more or less success different aspects of complex systems, but without encompassing theory behind them. I see this danger coming in particular from the remnants of reductionism and determinism that still influence many complexity researchers' way of thinking. Physicists especially have been trained to as much as possible make complete and deterministic models of the phenomena they study, albeit at the cost of studying only relatively simple aspects isolated from their environment or context. This allows them to make more accurate predictions than scientists in, say, biology, medicine or the social sciences, where the subject of investigation cannot be neatly separated out from the things it is connected to.

Now that physicists have started to focus on complexity they tend to take that same attitude with them, applying their impressive array of mathematical tools to the analysis of social, economical or biological systems. While this may produce plenty of interesting insights in the short term, in the long term they need to become aware that it will never provide them with the kind of absolutist "laws of complexity" that many still are looking for. Every complex system has followed its unique evolutionary trajectory and as such is different from any other system. It is only when we become deeply aware of the unlimited number of differences and connections between systems, and the unpredictable evolution this engenders that we will be able to truly build a science of complexity.

## Suggested reading

Ashby, W. R. (1962). Principles of the Self-organizing System. In von Foerster, H. and G. W. Zopf, Jr. (Eds.), *Principles of Self-organization*. Pergamon Press, pp. 255-278.

Gershenson C. & F. Heylighen (2004). How can we think the complex? in: Richardson, Kurt (ed.) Managing the Complex Vol. 1: Philosophy, Theory and Application.(Institute for the Study of Coherence and Emergence/Information Age Publishing)

Heylighen F. (1997): "Classic Publications on Complex, Evolving Systems: a citation-based survey, Complexity 2 (5), p. 31-36.

Heylighen F. (2001): "The Science of Self-organization and Adaptivity", in: L. D. Kiel, (ed.) Knowledge Management, Organizational Intelligence and Learning, and Complexity, in: The Encyclopedia of Life Support Systems ((EOLSS), (Eolss Publishers, Oxford). [http://www.eolss.net]

Heylighen F., P. Cilliers, & C. Gershenson (2007): "Complexity and Philosophy", in: Jan Bogg and Robert Geyer (editors), Complexity, Science and Society, (Radcliffe Publishing, Oxford)

Holland J.H. 1996 Hidden Order: How adaptation builds complexity, Addison-Wesley.

Kauffman S. A. 1995, At Home in the Universe: The Search for Laws of Self-Organization and Complexity, Oxford University Press, Oxford.

Prigogine, I. and Stengers, I. 1984 . Order out of Chaos, Bantam Books, New York,

Waldrop, M. M. (1992) Complexity: The Emerging Science at the Edge of Order and Chaos, London: Viking.